\documentclass[aps,floatfix,superscriptaddress]{revtex4}
\usepackage{graphicx,amssymb,amsfonts,amsmath,color,bm,hyperref,subfig}
\usepackage{xcolor}
\hypersetup{
    colorlinks,
    linkcolor={blue},
    citecolor={blue},
    urlcolor={black}
}

\begin{document}
\title{Entropy production and large deviation function for systems with microscopically irreversible transitions}

\author{Bappa Saha}
\affiliation{Department of Physics, Indian Institute of Technology,
Kanpur-208 016, India}

\author{Sutapa Mukherji}
\affiliation{Department of Protein Chemistry and Technology, 
Central Food Technological Research Institute, 
Mysore-570 020, India}

\date{\today}
\begin{abstract}
We obtain the large deviation function for entropy production of the medium and its distribution function for two-site totally asymmetric simple exclusion process (TASEP)  and three-state unicyclic network.  Since such systems are described through microscopic irreversible transitions, we obtain  time-dependent transition rates by sampling the states of these systems at a regular short time interval $\tau$. These transition rates are used to derive the large deviation function for the entropy production in the nonequilibrium steady state and its asymptotic distribution function. The shapes of the large deviation function and the distribution function depend on the value of the mean entropy production rate which has a non-trivial dependence on the particle injection and withdrawal rates in case of TASEP.  Further, it is argued that in case of a TASEP, 
 the distribution function tends to be like a Poisson distribution for smaller 
 values of particle injection and withdrawal 
rates.
\end{abstract}

\maketitle

\section{Introduction}
The entropy production of the surrounding medium of a system driven out of equilibrium is arguably the most convenient tool for quantifying the irreversiblity. In the long time limit when the steady state value of the entropy production arising due to the boundary contributions could be neglected, the probability distribution function  $P(\Delta S_m,t)$ for the change in the medium entropy $\Delta S_m$ at time $t$, is known to satisfy certain symmetry relation, generally known as the fluctuation theorem \cite{evans1,gcsymm1,gcsymm2,kurchan,lebowitz,harris}:
\begin{equation}
\frac{P(\Delta S_m,t)}{P(-\Delta S_m,t)}=\lim_{t\rightarrow\infty} e^{\Delta S_m}.
\end{equation}
This symmetry relation implies that the probability of observing the entropy annihilation over a long time interval becomes negligibly small and  can be viewed as the  nonequilibrium generalization of the second law of thermodynamics. The derivation of the distribution function in the asymptotic time limit is often performed by finding out the corresponding large deviation function(LDF) 
$I(\Delta S_m)$ which is related to the distribution function as, 
\begin{equation}
I(\Delta S_m)\equiv \lim_{t\rightarrow\infty}-\frac{1}{t}\ln P(\Delta S_m,t).\label{ldf_intro}
\end{equation}
 The LDF plays the role of the free energy functions in equilibrium systems 
 \cite{ellis,touchette} and in the case of nonequilibrium systems, its symmetry
  property, $I(-\Delta S_m)=I(\Delta S_m)+\Delta S_m/t$, validates the 
  fluctuation theorem or the Gallavotti-Cohen symmetry \cite{gcsymm1,gcsymm2,kurchan,lebowitz,harris}.

For a system described by the continuous time Markov dynamics, a microscopic transition from its configuration $i$ to $j$ with a transition rate, $\omega_{ij}$, causes the entropy of the surrounding medium to change \cite{lebowitz,harris,schn,jchemphys,seifert2,seifert3} by the amount $\Delta S_m=\ln \left(\frac{\omega_{ij}}{\omega_{ji}}\right)$, where $k_B$, the Boltzmann constant, is assumed to be unity.  Of the many studies on the entropy production of the medium for systems with microscopic reversibility \cite{seifert2,seifert3,imparato,rkpzia,barato,pleimling2}, some recent works on the properties of the LDF and its symmetry relation can be found in \cite{imparato,pleimling2} where the authors studied the asymptotic distributions of the entropy production by finding out the LDF using a generating function based approach. In reference \cite{pleimling2}, the authors obtained the LDF for partially asymmetric simple exclusion processes and reaction-diffusion processes with microscopic reversibility. The emergence of a kink-like feature in the LDF at zero entropy production was argued to be generic for such processes. This kink could be characterized by the 
average value of the medium entropy production rate, making the entropy production rate a good candidate 
for quantifying the irreversibility in nonequilibrium systems.

  While there is a wide applicability of the above  formula for finding  the 
 medium entropy production for  nonequilibrium systems having bi-directional
  transitions between its various states, this formula 
  cannot be applied directly when we encounter a system with irreversible microscopic 
  transitions  between its states \cite{zeraati}. Totally asymmetric simple exclusion 
  processes(TASEP) where particles move in only one direction respecting the exclusion 
  principle is one such example of systems with irreversible microscopic transition. 
  Other examples include enzymatic reaction networks modeled by Michaelis-Menten 
  scheme, directed percolation etc. In these systems when some of the transition 
  rates $\omega_{ij}$ become zero, this formula predicts an infinite 
  entropy production which  have not been observed in realistic situations 
  \cite{oliveira,takeuchi}. To address this shortcoming, ben-Avraham {\it et} {\it al.} 
  \cite{pleimling1} proposed a regularization scheme by sampling the states of 
  the system at small time interval and obtained the modified 
  transition probabilities. Using those transition probabilities, they 
  computed the medium entropy production rate and its LDF 
  for a three-state irreversible loop. In reference \cite{zeraati}, 
  the authors used  a slightly different method by introducing small backward
   transitions and determined the lower bound for the average 
   rate of medium entropy production originating from the 
   predominant irreversible transitions.

In this paper, we obtain the effective time-dependent transition rates for systems with irreversible transitions by allowing the systems to undergo all possible allowed transitions over small time interval $\tau$ and, then, deriving the probabilities of transition between any two states at the end of the time interval, $\tau$. The new transition rates  are identical to the 
original transition rates  for small  $\tau$. These transition rates allow us to 
extend straightforwardly the generating function based approach
 used earlier \cite{lebowitz,imparato} to find out the LDF for the 
 mean entropy production for irreversible systems. For a two-site TASEP 
 and a three-state irreversible system, we first obtain the transition 
 probabilities between any two 
arbitrary states by solving the governing Master equations. 
The transition rates are obtained after keeping the leading 
order terms in $\tau$ in  the Taylor expansion of the time-dependent 
probabilities  and then differentiating with respect to $\tau$. These new 
rates are used to obtain the average rate of medium entropy production and further 
to obtain the LDF for the entropy production  through 
a saddle point approximation. At zero entropy production, 
the LDFs for both the models show a kink  which  
 can be characterized by the average entropy production rate \cite{mehl, pleimling1,pleimling2,speck_t}.  Finally, the LDF is used to find  the distribution 
of the entropy production in the long time limit.  In the case of a two-site 
TASEP, the average rate of medium entropy production monotonically 
increases with the particle injection and withdrawal rates $\alpha$. 
 For large values of the medium entropy production rate, the  distribution function 
for the entropy production appears like a Gaussian distribution. 
For low entropy production rates, the distribution function turns out to be  a non-Gaussian one.  The features of  the LDF that are responsible for producing the non-Gaussian shape  of the entropy distribution function are strongly (exponentially) suppressed in the case of large entropy production rates.

The rest of the paper is organized as follows. In section \ref{sec:2}, we introduce
 the entropy production formalisms for Markov jump processes and a 
 practical way to apply those for systems having unidirectional transitions. 
 In section \ref{sec:3},  we first compute the rate of medium entropy production
  in the nonequilibrium steady state for two-site TASEP with equal injection and 
  withdrawal rates denoted by $\alpha$.  Next, we  obtain the LDF 
  as well as the distribution function for the medium entropy production for 
  different values of $\alpha$. In section \ref{sec:4}, using the same lines of approach, 
  we obtain the analytical form of the LDF for entropy 
  production for a three-state irreversible loop. The results are summarized 
  in section \ref{sec:5}. 

\section{Systems with irreversible transitions}\label{sec:2}
In this section, we first present a brief overview of these relations \cite{schn,lebowitz,harris,jchemphys,seifert2,seifert3,tome} valid for stochastic dynamics modeled as continuous time Markovian dynamics with finite, discrete configuration space. Next, 
we elucidate the feasibility of using the known entropy production formulae for a 
system having one or more microscopically irreversible transitions between its finite number of discrete states.  Due to the microscopic   irreversibility inherent in the system, some 
of the transition rates involved in the entropy production formulations are zero. 
Here, without introducing negligibly small backward transition rates  
as was done in  \cite{rkpzia,zeraati}, 
we obtain the new set of transition rates by sampling the states of the 
system at a small time interval $\tau$. In the limit $\tau\rightarrow 0$, these new transition rates approach their original values, thus making the computations of entropy production rate and its LDF more accurate.

\subsection{Entropy productions for  Markov jump processes}
We consider a continuous-time Markov jump process, for the time interval $0\le t\le t_f$, with finite number of states. The dynamical evolution of the probability  $P_i(t)$, that the system is found in state $i$,  is described by the Master equation
\begin{eqnarray}
 \frac{\partial}{\partial t}P_i(t) &=& \sum_{j\neq i}\left(\omega_{ji}(t)P_j(t)-\omega_{ij}(t)P_i(t)\right)\nonumber\\
 &=& \sum_j T_{ij}P_j , \label{1}
\end{eqnarray}
where $\omega_{ji}$ and $\omega_{ij}$ are the transition rates for the 
jump from state $j$ to $i$ and from state $i$ to $j$, respectively. Equation(\ref{1}) can be written in matrix form as,
\begin{equation}
\frac{\partial}{\partial t}\mathbf{P}(t)= \mathbf{T}\mathbf{P}(t), \label{matrix_form}
\end{equation}
where $\mathbf{P}=(P_1, P_2,\dots)^T$ is the column matrix and the  $\{i,j\}$th element of the transition matrix $\mathbf{T}$  are
\begin{equation}
 T_{ij}=\omega_{ji}-\delta_{ij}\sum_{k\neq i} \omega_{ik}. \label{2}
\end{equation}

In order to obtain the expression for the entropy production due to a  transition 
 from one state to another, we begin by defining the average Gibbs entropy of the 
 system as
\begin{equation}
   \langle S\rangle  =  -\sum_i P_i \ln P_i.
   \end{equation}
The expression for the time evolution of the system entropy has the form \citep{schn,lebowitz,harris,jchemphys,seifert2,seifert3,tome}
\begin{eqnarray}
\langle \dot{S}\rangle = \sum_{i,j} P_i \omega_{ij}\ln\left(\frac{P_i\omega_{ij}}{P_j\omega_{ji}}\right)  
              -\sum_{i,j} P_i \omega_{ij}\ln\left(\frac{\omega_{ij}}{\omega_{ji}}\right), \label{12}
\end{eqnarray}
where the overdot implies a derivative with respect to time.
The first term on the right hand side of  the above relation is always positive 
and is identified as the total entropy production rate due to stochastic transitions.  
The  second term  is the medium entropy production rate or the entropy 
flow into the medium due to these transitions.
The total rate of the entropy production is now expressed as
\begin{equation}
  \langle\dot{S}_{tot}\rangle = \langle\dot{S}\rangle + \langle\dot{S}_m\rangle,
  \end{equation}
  with 
\begin{eqnarray}
  \langle\dot{S}_{tot}\rangle = \sum_{i,j}P_i\omega_{ij}\ln\left(\frac{P_i\omega_{ij}}{P_j\omega_{ji}}\right) \ \ {\rm and}\ \ 
   \langle\dot{S}_m\rangle = \sum_{i,j}P_i\omega_{ij}\ln\left( \frac{\omega_{ij}}{\omega_{ji}}\right). \label{medium_entropy}
  \end{eqnarray}

\subsection{Computations of time-dependent transition rates}
In order to calculate  the medium entropy production rate, we first 
briefly describe the strategy to compute the transition rates employing 
a matrix-based approach. To begin with, we consider a Markov process 
in which transitions between the discrete states $i=(1,2,\dots N)$ are 
measured in discrete time steps, $t=n\Delta t$, where $n=0,1,2\dots$. The solution of equation(\ref{matrix_form}) 
is determined by diagonalizing the matrix $\mathbf{T}$ as $\mathbf{T_D}=\mathbf{B}^{-1}\cdot \mathbf{T}\cdot \mathbf{B}$, where matrix $\mathbf{B}$ 
 is formed of the eigenvectors of $\mathbf{T}$ arranged column-wise, and 
 $\mathbf{B}^{-1}$ is the inverse of $\mathbf{B}$.  The diagonal matrix $\mathbf{T_D}$ 
 has the eigenvalues of $\mathbf{T}$ as its  diagonal elements. 
 The solution of 
 equation(\ref{matrix_form}) then reads
\begin{eqnarray}
\mathbf{P}(t)=\mathbf{B}\cdot e^{t\mathbf{T_D}}\cdot \mathbf{B}^{-1}\cdot \mathbf{P}(0)=\mathbf{\bar{T}}\cdot \mathbf{P}(0).\label{soln1}
\end{eqnarray}   
The elements of the $\mathbf{\bar{T}}$ matrix are the transition 
probabilities. For instance, the element $\bar{T}_{ij}$, in conventional 
notation, implies $\bar{T}(i,t|j,0)$, i.e., the probability of finding the system 
at $i$-th state at time $t$, provided it was at $j$-th state at the initial time. 
These transition probabilities allow us to obtain the time-dependent 
transition rates \cite{reichl} if the states of the system are sampled at 
small time interval $\Delta t$. To be more specific, let us consider the
 probability $P_i(t+\Delta t)$ at time $t+\Delta t$ of finding the system at state $i$. 
 In the limit  $\Delta t\rightarrow 0$, we have,
\begin{eqnarray}
P_i(t+\Delta t)\approx P_i(t)+\Delta t\frac{\partial}{\partial t}P_i(t)\nonumber\\
= P_i(t)+\Delta t \sum_j W_{ij}P_j(0). \label{wmatrix}
\end{eqnarray}
This relation defines   $W_{ij}$  which denotes 
 the transition rate from state $j$ to $i$. In the subsequent sections, these 
 transition rates are used in relation (\ref{medium_entropy}) to calculate the average 
 rate of entropy production of the medium, its LDF and the distribution function 
 in the asymptotic time limit.

\section{Entropy production for two-site TASEP}\label{sec:3}
\subsection{Time-dependent transition rates}
\begin{figure}[ht!]
  \centering
   \includegraphics[height=.33\textwidth]{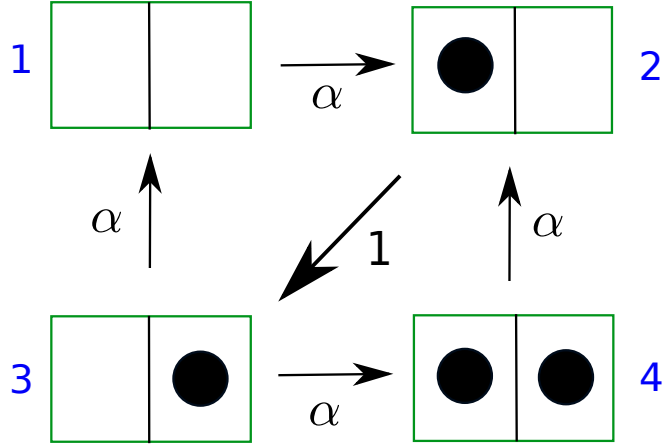}  
\caption{Four states with the transition rates for  two-site TASEP.}
\label{fig:samplefig1}
\end{figure}

We consider  two-site TASEP with equal particle injection and withdrawal rates $\alpha$ as our first  model of a system with irreversible transitions between its four states as shown in figure \ref{fig:samplefig1}.
Let us consider the time evolution of the probability densities  $\mathbf{P}=(P_1,P_2,P_3,P_4)^T$, where the element $P_i$ of the column matrix denotes the probability of finding the system in state $i$. The governing Master equation can be written as,
\begin{equation}
\frac{\partial}{\partial t}\mathbf{P}(t)=\mathbf{T}\mathbf{P}(t), \label{firsteq}
\end{equation}
where $\mathbf{T}$ is the $4\times 4$ matrix having the following form,
\begin{equation}
\mathbf{T} = \left( \begin{array}{rrrr}
                                   -\alpha & 0 & \alpha & 0\\
                                    \alpha & -1 & 0 & \alpha\\
                                    0 & 1 & -2\alpha & 0\\
                                    0 & 0 & \alpha & -\alpha\\
\end{array}\right). \label{tasep_trmatrix}
\end{equation}

The eigenvalues of $\mathbf{T}$ are  $\lambda_1=0,\ \lambda_2=-\alpha,\ \lambda_3=-\frac{1}{2}[1+3\alpha+\sqrt{1-6\alpha+\alpha^2}],\ \lambda_4=\frac{1}{2}[-1-3\alpha+\sqrt{1-6\alpha+\alpha^2}]$  with the corresponding  eigenvectors 
\begin{eqnarray}
e_1=(1, 2\alpha, 1, 1)^T;\  e_2=(-1,0,0,1)^T;\  e_3= \left(1,\frac{1}{2\alpha}[1-3\alpha+\sqrt{1-6\alpha+\alpha^2}],-\frac{1}{2\alpha}[1+\alpha +\sqrt{1-6\alpha +\alpha^2}],1\right)^T;\\
e_4=\left(1,\frac{1}{2\alpha}[1-3\alpha -\sqrt{1-6\alpha +\alpha^2}],-\frac{1}{2\alpha}[1+\alpha-\sqrt{1-6\alpha+\alpha^2}],1\right)^T.
\end{eqnarray}
The matrix $\mathbf{B}^{-1}$ has the form 
\begin{eqnarray}
\mathbf{B}^{-1}=\frac{1}{3+2\alpha} \left( \begin{array}{cccc}
                                   1 & 1 & 1 & 1\\
                                    -(3+2\alpha)/2 & 0 & 0 & (3+2\alpha)/2\\
                                    \frac{\alpha-1-2\alpha^2 +m(1+2\alpha)}{4m} & \frac{1+3\alpha -m}{2m} & \frac{1-\alpha(3+4\alpha)-m}{2m} & \frac{\alpha-1 -2\alpha^2+ m(1+2\alpha)}{4m}\\
                                    \frac{1-\alpha+2\alpha^2 +m(1+2\alpha)}{4m} & \frac{-(1+3\alpha+m)}{2m} & \frac{\alpha(3+4\alpha)-1-m}{2m} & \frac{1-\alpha+2\alpha^2 +m(1+2\alpha)}{4m}\\
\end{array}\right),
\end{eqnarray}
with  $m=\sqrt{1+\alpha(\alpha-6)}$. 

Substituting $\mathbf{B}$, $\mathbf{T}_D$ and $\mathbf{B}^{-1}$ in equation(\ref{soln1}), each element  of the column vector ${\bf P}(t)$  is expressed as 
\begin{eqnarray}
P_i(t)=\sum_j \bar{T}_{ij} P_j(0),
\end{eqnarray}
which can be written in the compact form as 
\begin{equation}
\mathbf{P}(t)=\mathbf{\bar{T}}\mathbf{P}(0). \label{psoln1}
\end{equation}
In the above equation, the conditional probability $\bar{T}_{ij}$ implies the probability of finding the system at $i$-th state at time $t$, provided it was at $j$-th state at initial time. $\bar{T}_{ii}$ term corresponds to the null transition. If we consider the time interval, $t=\tau$, to be small such that the sampling time becomes, $\tau<<1/\alpha$, it is then ensured that the transition matrix $\mathbf{W}$ becomes closer to the original transition matrix (\ref{tasep_trmatrix}). In this limit, the matrix $\mathbf{\bar{T}}$ has the form,
\begin{flalign}
\mathbf{\bar{T}}= \left( \begin{array}{cccc}
                                   1-\alpha \tau+ \frac{\alpha^2\tau^2}{2} & \frac{\alpha \tau^2}{2} & \alpha\tau -\frac{3\alpha^2\tau^2}{2} & \frac{\alpha^2\tau^3}{6}\\
                                    \alpha\tau- \frac{\alpha(1+\alpha)\tau^2}{2} & 1-\tau+\frac{\tau^2}{2} & \alpha^2\tau^2 & \alpha\tau- \frac{\alpha(1+\alpha)\tau^2}{2} \\
                                    \frac{\alpha \tau^2}{2} & \tau-(\alpha+ \frac{1}{2})\tau^2 & 1-2\alpha\tau +2\alpha^2\tau^2 & \frac{\alpha\tau^2}{2} \\
                                    \frac{\alpha^2 \tau^3}{6} & \frac{\alpha \tau^2}{2} & \alpha \tau- \frac{3\alpha^2\tau^2}{2} & 1-\alpha\tau+\frac{\alpha^2\tau^2}{2}\\
\end{array}\right).
\end{flalign}
 The corresponding transition matrix $\mathbf{W}$ as defined in (\ref{wmatrix}) is
 \begin{flalign}
\mathbf{W}= \left( \begin{array}{cccc}
                                   \alpha^2\tau -\alpha & \alpha\tau & \alpha-3\alpha^2\tau & \frac{\alpha^2\tau^2}{2}\\
                                    \alpha-\alpha(1+\alpha)\tau & \tau-1 & 2\alpha^2\tau & \alpha-\alpha(1+\alpha)\tau \\
                                    \alpha\tau & 1-2(\alpha+\frac{1}{2})\tau & 4\alpha^2\tau-2\alpha & \alpha\tau \\
                                    \frac{\alpha^2\tau^2}{2} & \alpha\tau & \alpha-3\alpha^2\tau & \alpha^2\tau- \alpha\\
\end{array}\right).
\end{flalign}

\subsection{Average entropy production rate of the medium} 
 Having obtained the time-dependent transition rates in the previous 
 subsection, we now evaluate the average rate of medium entropy
  production in the nonequilibrium steady state as defined in 
  equation(\ref{medium_entropy}). We define the transition rate for a 
  transition from state $j$ to $i$ as $\tilde{\omega}_{ji}$ which is 
  related to the corresponding element of the transition 
  matrix as, $\tilde{\omega}_{ji}=W_{ij}$. The steady state probability 
  densities for the two-site model are obtained as,
 \begin{eqnarray}
 P_{1s}= P_{3s}=P_{4s}=\frac{1}{3+2\alpha};\ \ \ \ P_{2s}=\frac{2\alpha}{3+2\alpha}.
 \end{eqnarray}
In the small time limit, the ratio of the forward and the reverse
 transition rates are approximated as,
\begin{eqnarray}
\frac{\tilde{\omega}_{12}}{\tilde{\omega}_{21}}\approx \frac{\tilde{\omega}_{31}}{\tilde{\omega}_{13}} \approx \frac{\tilde{\omega}_{42}}{\tilde{\omega}_{24}}\approx \frac{\tilde{\omega}_{34}}{\tilde{\omega}_{43}} \approx \frac{1}{\tau},\\
\frac{\tilde{\omega}_{23}}{\tilde{\omega}_{32}} \approx \frac{1}{2\alpha^2\tau}.
\end{eqnarray}
The average rate of  entropy production of the medium is thus obtained as,
\begin{eqnarray}
\langle\dot{S}_m\rangle &=& \sum_{i,j}P_{is}\tilde{\omega}_{ij}\ln \frac{\tilde{\omega}_{ij}}{\tilde{\omega}_{ji}}\nonumber\\
&=&\frac{2\alpha(1-\tau-3\alpha\tau)}{(3+2\alpha)}\left[2\ln(1/\tau)+\ln(1/(2\alpha^2\tau))\right].
\end{eqnarray}
The entropy production rate is plotted in the
 figure \ref{fig:entropy_prodiction_eigenvalues}(a) with $\alpha$. The positivity of the entropy production suggests that the medium  entropy increases as the system undergoes transition from one state to another.
 
\subsection{Large deviation function for entropy production and its distribution function}
With the definition of the transition matrix $\mathbf{W}$ in the previous section, 
we calculate the LDF for entropy production\cite{lebowitz,imparato}. 
 Let $\phi_i(\Delta S_m,t)$ be the probability that the system is in the $i$-th 
 state at time $t$ while the change in the medium entropy is $\Delta S_m$. 
 The probability of finding the system at time $t+\tau$ after a small 
 time interval $\tau$, during which the entropy exchange with 
 the medium is $\Delta s_{ji}=\ln\left(\frac{\tilde{\omega}_{ji}}{\tilde{\omega}_{ij}}\right)$ due to the jump of the system from state $j$ to $i$, is expressed as \cite{imparato},
\begin{eqnarray}
\phi_i(\Delta S_m, t+\tau)\approx \phi_i(\Delta S_m, t)+ 
\tau\sum_j \left[\tilde{\omega}_{ji}\phi_j(\Delta S_m-\Delta s_{ji},t)-
\tilde{\omega}_{ij}\phi_i(\Delta S_m,t)\right].
\end{eqnarray}
In the limit $\tau\rightarrow 0$, we have,
\begin{eqnarray}
\frac{\partial \phi_i}{\partial t}= \sum_j \left[ \tilde{\omega}_{ji}\left(\sum_{n=0}^{\infty}\frac{(-\Delta s_{ji})^n}{n!}\frac{\partial ^n}{\partial (\Delta S_m)^n}\phi_j\right) -\tilde{\omega}_{ij}\phi_i\right].
\end{eqnarray}
Introducing the generating function 
\begin{eqnarray}
\psi_i(\lambda,t)=\int d(\Delta S_m) e^{-\lambda \Delta S_m}\phi_i, \label{genfn}
\end{eqnarray} 
we write the time evolution of the generating function as,
\begin{eqnarray}
\frac{\partial \psi_i}{\partial t}&=& \sum_j \tilde{\omega}_{ji}\psi_j e^{-\lambda \Delta s_{ji}}-\sum_j \tilde{\omega}_{ij}\psi_i\nonumber \\
&=&\sum_j \tilde{\omega}_{ji}^{1-\lambda}\tilde{\omega}_{ij}^\lambda\psi_j -\sum_j \tilde{\omega}_{ij} \psi_i
=\sum_j L_{ij}\psi_j .\label{generating_fn}
\end{eqnarray}
The above equation can be written in a  matrix form as,
\begin{eqnarray}
\frac{\partial {\boldsymbol{\psi}}}{\partial t} = \mathbf{L}{\boldsymbol{\psi}},
\end{eqnarray}
where 
\begin{flalign}
&\mathbf{L}= \left( \begin{array}{llll}
                                   -(\alpha +\alpha^2\tau(\tau-2)/2) & (\alpha\tau)^{1-\lambda} (\alpha-\alpha(1+\alpha)\tau)^{\lambda} & (\alpha-3\alpha^2\tau)^{1-\lambda}(\alpha\tau)^{\lambda} & \frac{\alpha^2\tau^2}{2}\\
                                    (\alpha-\alpha(1+\alpha)\tau)^{1-\lambda}(\alpha\tau)^{\lambda} & \tau-1 & (2\alpha^2\tau)^{1-\lambda} [1-2(\alpha+0.5)\tau]^{\lambda} & (\alpha-\alpha(1+\alpha)\tau)^{1-\lambda}(\alpha\tau)^{\lambda} \\
                                    (\alpha\tau)^{1-\lambda}(\alpha-3\alpha^2\tau)^{\lambda} & (1-2(\alpha+0.5)\tau)^{1-\lambda}(2\alpha^2\tau)^{\lambda} & (4\alpha^2\tau-2\alpha) & (\alpha\tau)^{1-\lambda}(\alpha-3\alpha^2\tau)^{\lambda} \\
                                    \frac{\alpha^2\tau^2}{2} & (\alpha\tau)^{1-\lambda}(\alpha-\alpha(1+\alpha)\tau)^{\lambda} & (\alpha-3\alpha^2\tau)^{1-\lambda}(\alpha\tau)^{\lambda} & -(\alpha+\alpha^2\tau(\tau-2)/2)\\
\end{array}\right),&
\end{flalign}
and ${\boldsymbol\psi}=(\psi_1,\psi_2,\psi_3,\psi_4)^T$.

\begin{figure}[htbp]
  \begin{center}
  (a) \includegraphics[width=.45\textwidth,clip,
angle=0]{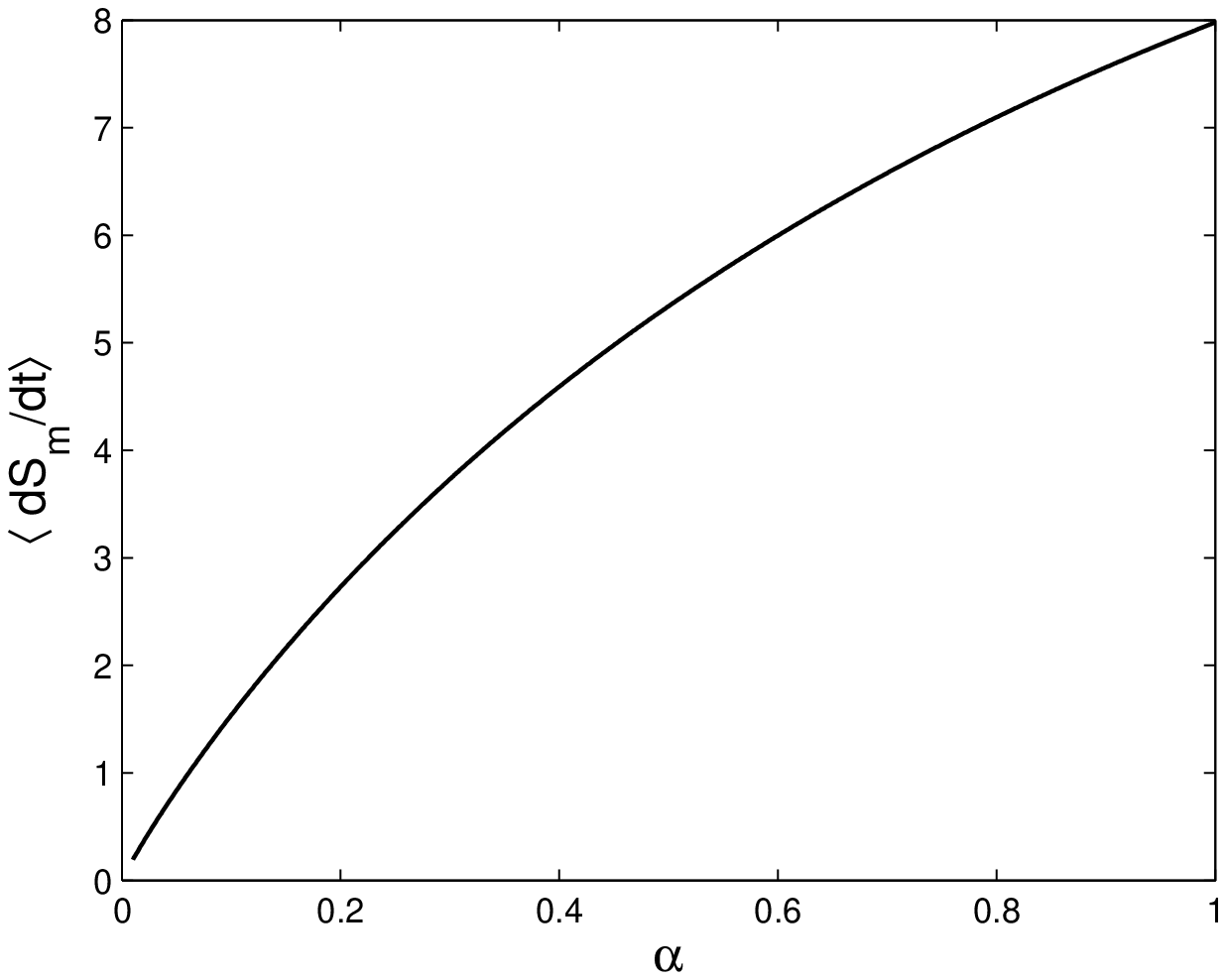}
(b)   \includegraphics[width=.45\textwidth,clip,
angle=0]{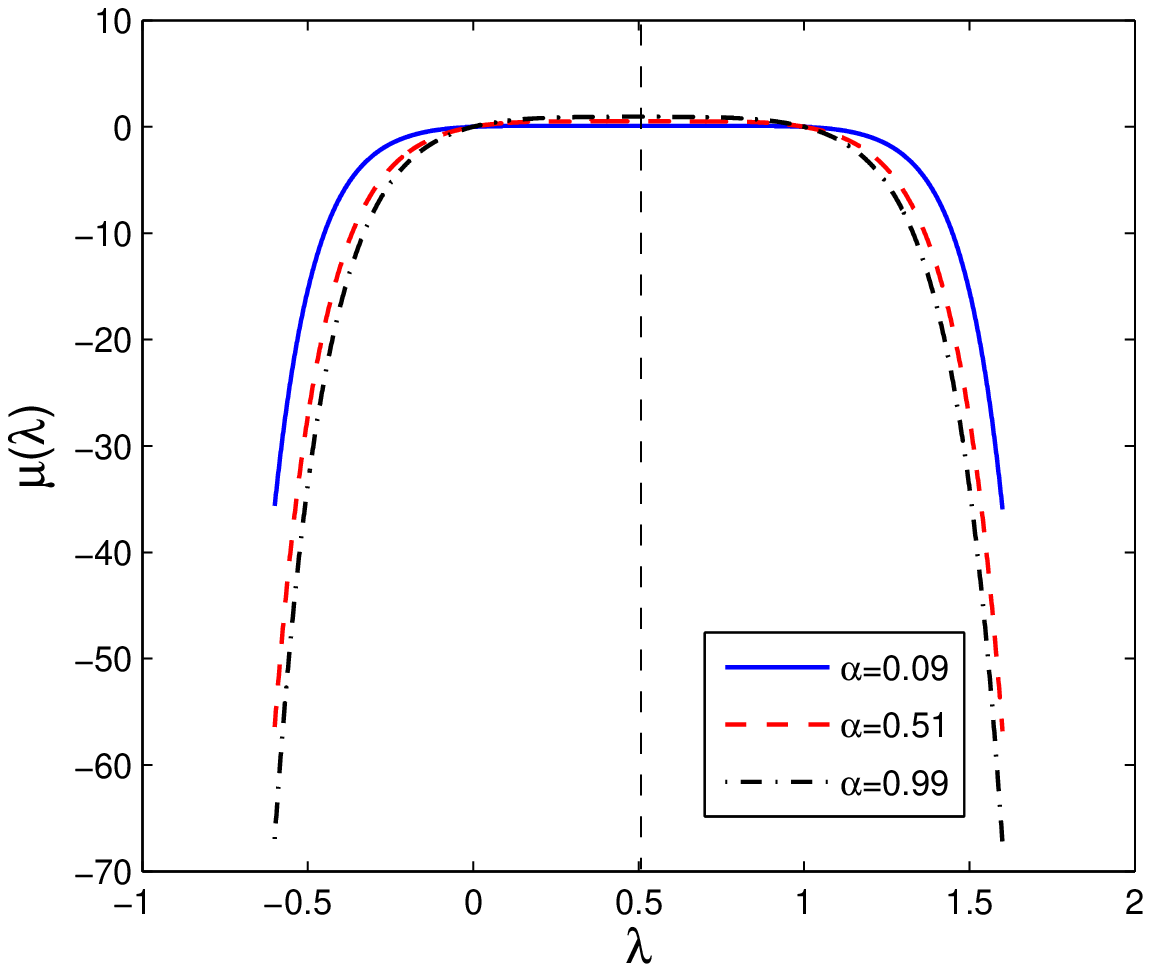}
\caption{(a) Rate of average entropy production of the medium. The value of the small time
 interval, $\tau=0.001$. (b) The smallest eigenvalue, $\mu(\lambda)$, for different values 
 of particle injection/withdrawal rates $\alpha$. For smaller values of $\alpha$, it becomes 
 more flat around $\mu=0$. The symmetry about the vertical dashed line at 
 $\lambda=0.5$ is the manifestation of the fluctuation theorem.}
\label{fig:entropy_prodiction_eigenvalues}
  \end{center}
\end{figure}

\begin{figure}[ht!]
 \begin{center}
 (a)  \includegraphics[width=.45\textwidth,clip,
angle=0]{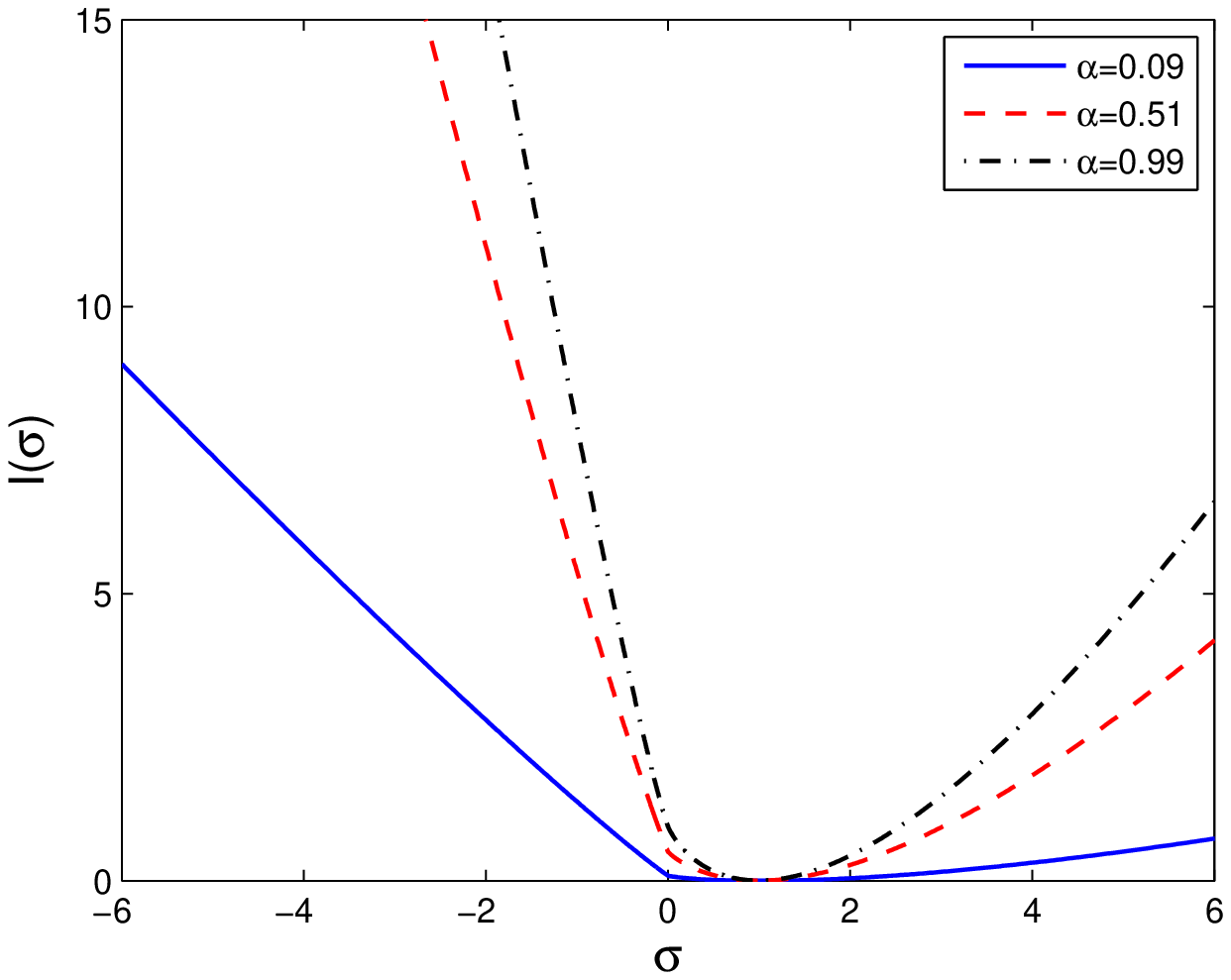} 
 (b)   \includegraphics[width=.45\textwidth,clip,
angle=0]{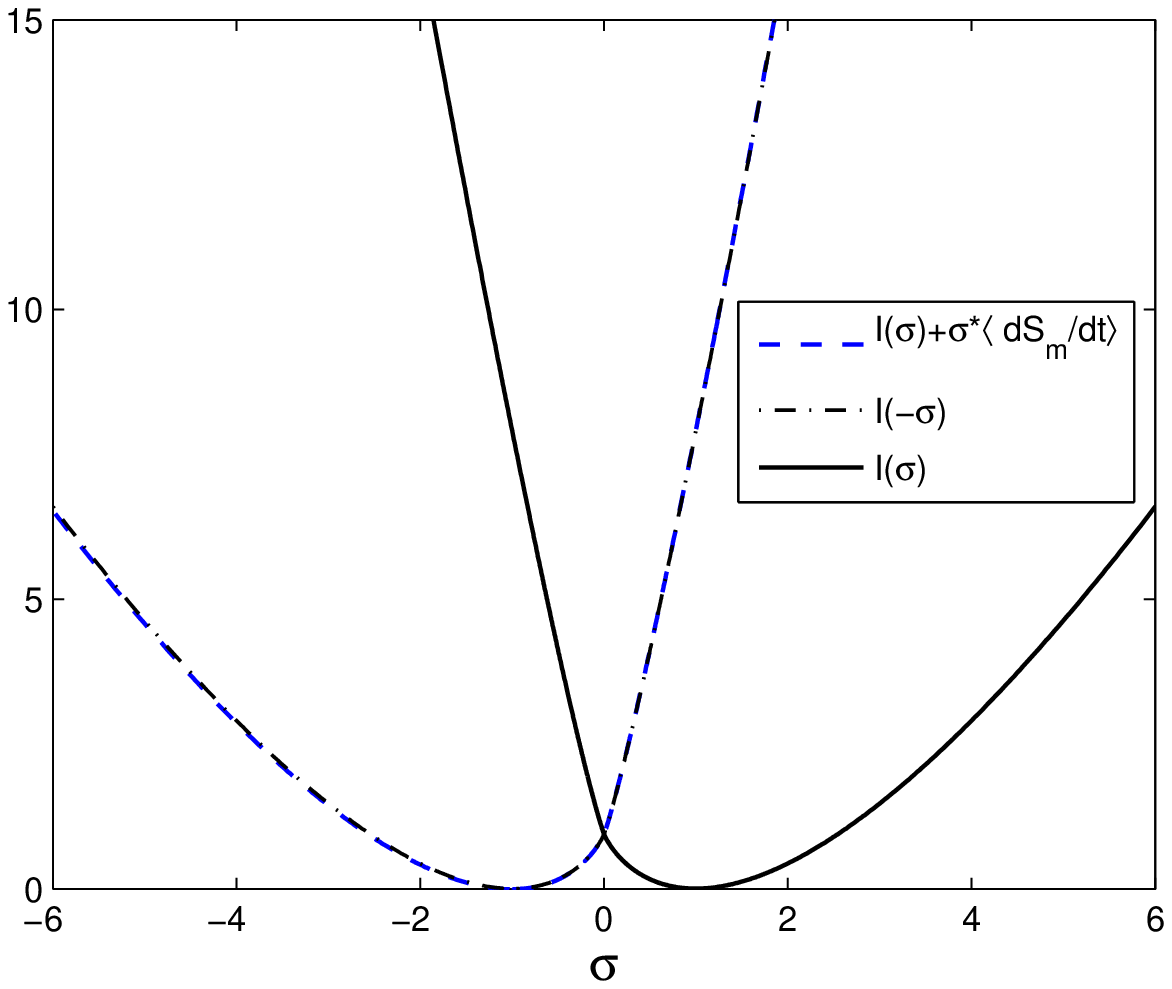} 
\caption{(a) Large deviation function for the scaled entropy
 production $\sigma$ for different values of particle injection and withdrawal 
 rates, $\alpha$. The kink like feature in the LDF at zero entropy production 
 becomes more prominent for larger values of $\alpha$. 
The blue solid line, the red dashed line and the black 
dashed-dot line correspond to the values of the average rate of entropy production $1.379, 5.389, \textrm{and}\ 7.911$ respectively. (b) This figure displays the 
validity of the symmetry relation, $I(\sigma)+\langle\dot{S}_m\rangle\sigma =I(-\sigma)$,
satisfied by  the LDF. The dashed line and the dashed-dot line have merged 
completely in the figure.}
\label{fig:largedeviation}
\end{center}
\end{figure}

 In order to find  the  total probability distribution 
$\phi(\Delta S_m,t)=\sum_i \phi_i(\Delta S_m,t)$, we need to introduce
 the total generating function $\psi(\lambda,t)=\sum_i \psi_i(\lambda,t)$.
In the large time limit, the total generating function can be approximated as, 
\begin{equation}
\psi(\lambda) \approx \lim_{t\rightarrow \infty} \exp(-t \mu_0(\lambda)),
\end{equation}
where $\mu_0(\lambda)$ is the smallest  of all the eigenvalues defined through 
the following equation
\begin{eqnarray}
\mathbf{L} \mathbf{l_n}(\lambda)= - \mu_n(\lambda) \mathbf{l_n}(\lambda).
\end{eqnarray}
We evaluate the smallest eigenvalue numerically and it is plotted in figure \ref{fig:entropy_prodiction_eigenvalues}(b) with $\lambda$. The 
symmetry of the smallest eigenvalue about $\lambda=0.5$ validates 
the fluctuation theorem $\mu_0(\lambda)=\mu_0(1-\lambda)$ \cite{gcsymm1,gcsymm2,kurchan,lebowitz,harris}.
The average rate of entropy production of the medium $\langle\dot{S}_m\rangle$ is related to the smallest eigenvalue as,
\begin{equation}
\left\langle \dot{S}_m\right\rangle = \frac{\partial \mu_0(\lambda)}{\partial \lambda}|_{\lambda=0}.
\end{equation}

In order to obtain the probability 
distribution $\phi_i(\Delta S_m,t)$, one has to  invert the relation 
in equation (\ref{genfn}).  The final integration 
is done using a saddle point approximation scheme.
 The LDF or the rate function $I(\sigma)$ with  
$\sigma= \frac{\Delta S_m}{t\left\langle \dot{S}_m\right\rangle}$ as 
the normalised rate of entropy production,   can be expressed 
as the Legendre transform of $\mu_0(\lambda)$ 
  \begin{equation}
 I(\sigma)=\mu_0(\lambda^*)-\lambda^* \sigma\left\langle \dot{S}_m\right\rangle.
 \end{equation}

Here $\lambda^*$ is the saddle point defined through the equation 
$\frac{\partial \mu_0(\lambda)}{\partial \lambda}|_{\lambda^*}=\sigma\left\langle \dot{S}_m\right\rangle$. From figure \ref{fig:entropy_prodiction_eigenvalues}(a), it is evident that the average 
entropy production rate is always positive and it increases as we 
increase the value of $\alpha$. At  zero 
entropy production, the LDF shows a kink  which becomes prominent for larger 
values of $\alpha$ (see figure \ref{fig:largedeviation}(a)). The 
symmetry property displayed by the 
LDF, $I(\sigma)+\langle\dot{S}_m\rangle\sigma =I(-\sigma)$, as shown in  figure \ref{fig:largedeviation}(b), is attributed to the  symmetry property 
of the distribution function of entropy production quantified through the fluctuation theorem.   

\begin{figure}[t!]
  \begin{center}
   \includegraphics[width=1\textwidth, clip, angle=0]{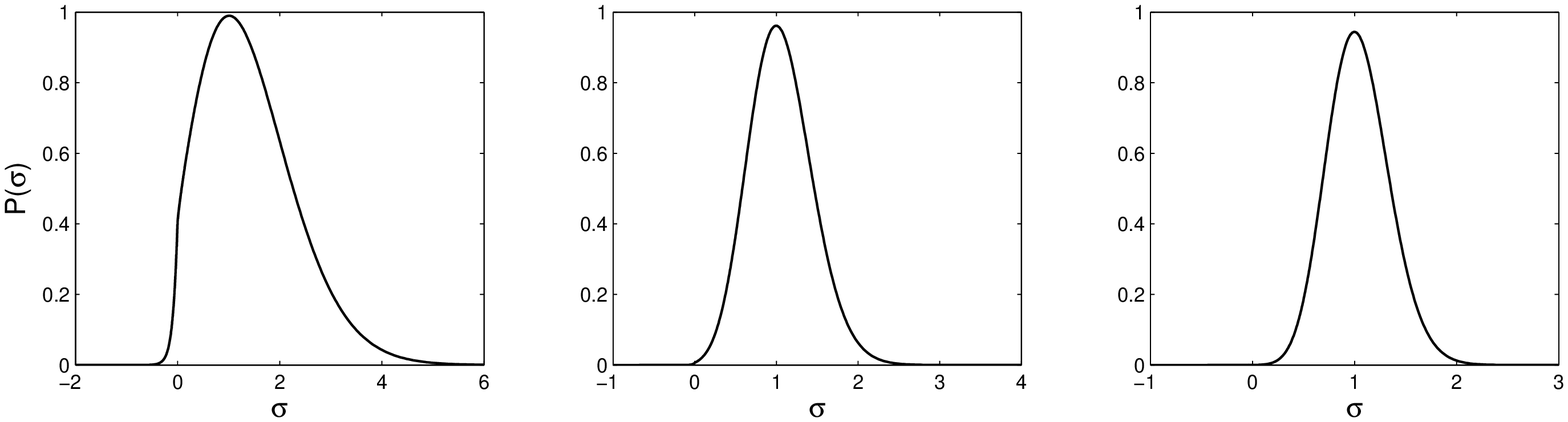}  
\caption{Probability distribution for the scaled entropy production at time $t=10$. From the left to right panel, the three figures correspond to  the rates  $\alpha=0.09$, $\alpha=0.51$ and $\alpha=0.99$, respectively. For the small values of the entry and exit rates, the distribution is non-Gaussian.  The distribution tends to be Gaussian for larger values of entry 
and exit  rates.}
\label{fig:entropy_prodiction_dist}
\end{center}
\end{figure}

\begin{figure}[h!]
  \begin{center}
   \includegraphics[width=.46\textwidth, clip, angle=0]{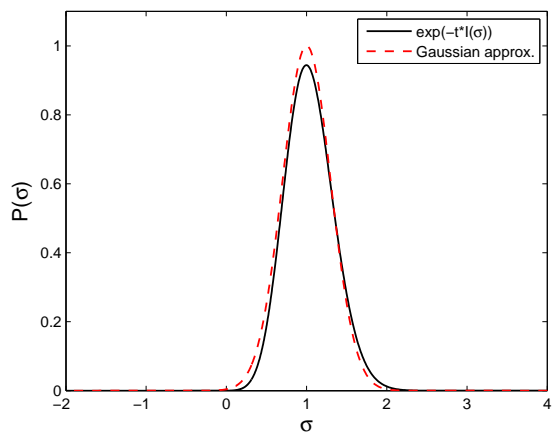}  
\caption{Probability distribution for the scaled entropy production at time $t=10$ for 
$\alpha=0.99$. For this value of $\alpha$, $\langle\dot{S}_m\rangle=7.911$.
The solid black curve corresponds to the $\exp(-t*I(\sigma))$ and the red dashed curve 
is obtained using equation(\ref{gaussian_approx}). }
\label{fig:gaussian_approx}
\end{center}
\end{figure}

Using the  expression for the LDF,  one may find out the 
 probability distribution function of the normalized entropy production rate $P(\sigma)$ 
 in the asymptotic time limit as 
\begin{equation}
P(\sigma)\sim \lim_{t\rightarrow\infty} \exp(-tI(\sigma)).
\end{equation}
The LDF provides the detailed information about the distribution function, 
which is non-Gaussian in nature in our case, for large fluctuations.
However, for larger values of $\alpha$ when the value of  the average entropy 
production rate is also large, the central part of the distribution tends
 to be Gaussian, while for smaller values of $\alpha$ it has a 
 non-Gaussian form. Intuitively, for small $\alpha$, the number of 
 transitions are also small and over the time interval $t=10$, the 
 smaller number of events cause the distribution function to have a Poisson 
 distribution-like form. The distribution functions $P(\sigma)$ 
 for three different values of $\alpha$ are 
 plotted in figure \ref{fig:entropy_prodiction_dist}.

To have a qualitative understanding of the distribution function 
for larger values of $\langle\dot{S}_m\rangle$, it should be noted 
from figure \ref{fig:largedeviation}(a) that the values of the LDF 
from its minimum are larger if the rate of medium entropy 
production is increased. These large fluctuations are strongly 
suppressed in the exponential form of the distribution function. 
Thus, if we perform a Taylor expansion of the LDF about its 
minimum at $\sigma=1$, we have
\begin{equation}
I(\sigma)\approx \frac{1}{2} \frac{d^2I(\sigma)}{d\sigma^2}|_{\sigma=1} (\sigma-1)^2.\label{gaussian_approx}
\end{equation}
We determine the second derivative numerically and 
for $\langle\dot{S}_m\rangle =7.911$,  its value is, 
$\frac{d^2I(\sigma)}{d\sigma^2}|_{\sigma=1}=1.101$. Taking 
the form of $I(\sigma)$ as in the equation(\ref{gaussian_approx}), 
we obtain the distribution function for large values of medium 
entropy production (see figure \ref{fig:gaussian_approx}). The 
matching between the original distribution and the approximated 
one is remarkable and thus, the distribution can be approximated as
 Gaussian for large $\langle\dot{S}_m\rangle$. However, similar 
 approximation cannot be made for smaller values of $\langle\dot{S}_m\rangle$ 
 since in this case, the fluctuations away from the center are not so large. 
 This  explains why the distribution function in this case becomes non-Gaussian.

\section{Three-state unicyclic network}\label{sec:4}

Here  we apply the present method to a three-state
 irreversible loop \cite{pleimling1} where 
the transitions between the three states denoted  as $1,\  2\ {\rm and} \ 3$ happen in 
a cyclic way as $1\rightarrow 2,\ 2\rightarrow 3,\ 3\rightarrow 1$ with rate $1$. 
In the small time interval $\tau$, the  transition rates $\tilde{\omega}_{ij}$s have the form,
\begin{eqnarray}
\tilde{\omega}_{12}=\tilde{\omega}_{23}=\tilde{\omega}_{31}\approx 1-2\tau, \\
\tilde{\omega}_{21}=\tilde{\omega}_{32}=\tilde{\omega}_{13}\approx \tau .
\end{eqnarray}
As before, the time evolution of the generating function $\psi_i$ $(i=1,2,3)$, as defined in equation(\ref{genfn}),  is governed by,
\begin{equation}
\frac{\partial {\boldsymbol{\psi}}}{\partial t}= \mathbf{L}{\boldsymbol{\psi}} ,
\end{equation}
where ${\boldsymbol{\psi}}=(\psi_1, \psi_2, \psi_3)^T$ and  $\mathbf{L}$ has the form,
\begin{eqnarray}
\mathbf{L}=\left( \begin{array}{lll}
                                   \tau -1 & (1-2\tau)^{\lambda}\tau^{1-\lambda} & (1-2\tau)^{1-\lambda}\tau^{\lambda} \\
                                    (1-2\tau)^{1-\lambda}\tau^{\lambda} & \tau-1 & (1-2\tau)^{\lambda}\tau^{1-\lambda} \\
                                    (1-2\tau)^{\lambda}\tau^{1-\lambda} & (1-2\tau)^{1-\lambda}\tau^{\lambda} & \tau-1 \\
\end{array}\right).
\end{eqnarray}
The smallest eigenvalue of ${\bf L}$  dominates the large time behavior of the total generating function $\psi=\sum_i\psi_i$. In this case, the smallest eigenvalue is 
\begin{eqnarray}
\mu_0(\lambda)= 1-\tau -(1-2\tau)^{\lambda}\tau^{1-\lambda}-(1-2\tau)^{1-\lambda}\tau^{\lambda}. \label{smallesteigenv}
\end{eqnarray}
The rate of medium entropy production is found as,
\begin{equation}
\langle \dot{S}_m\rangle = \frac{\partial \mu_0 }{\partial \lambda}|_{\lambda=0} = (1-3\tau)\ln[(1-2\tau)/\tau].\label{med_entropy_pleimling}
\end{equation}
\begin{figure}[ht!]
  \centering 
   (a) \includegraphics[width=.45\textwidth,clip,
angle=0]{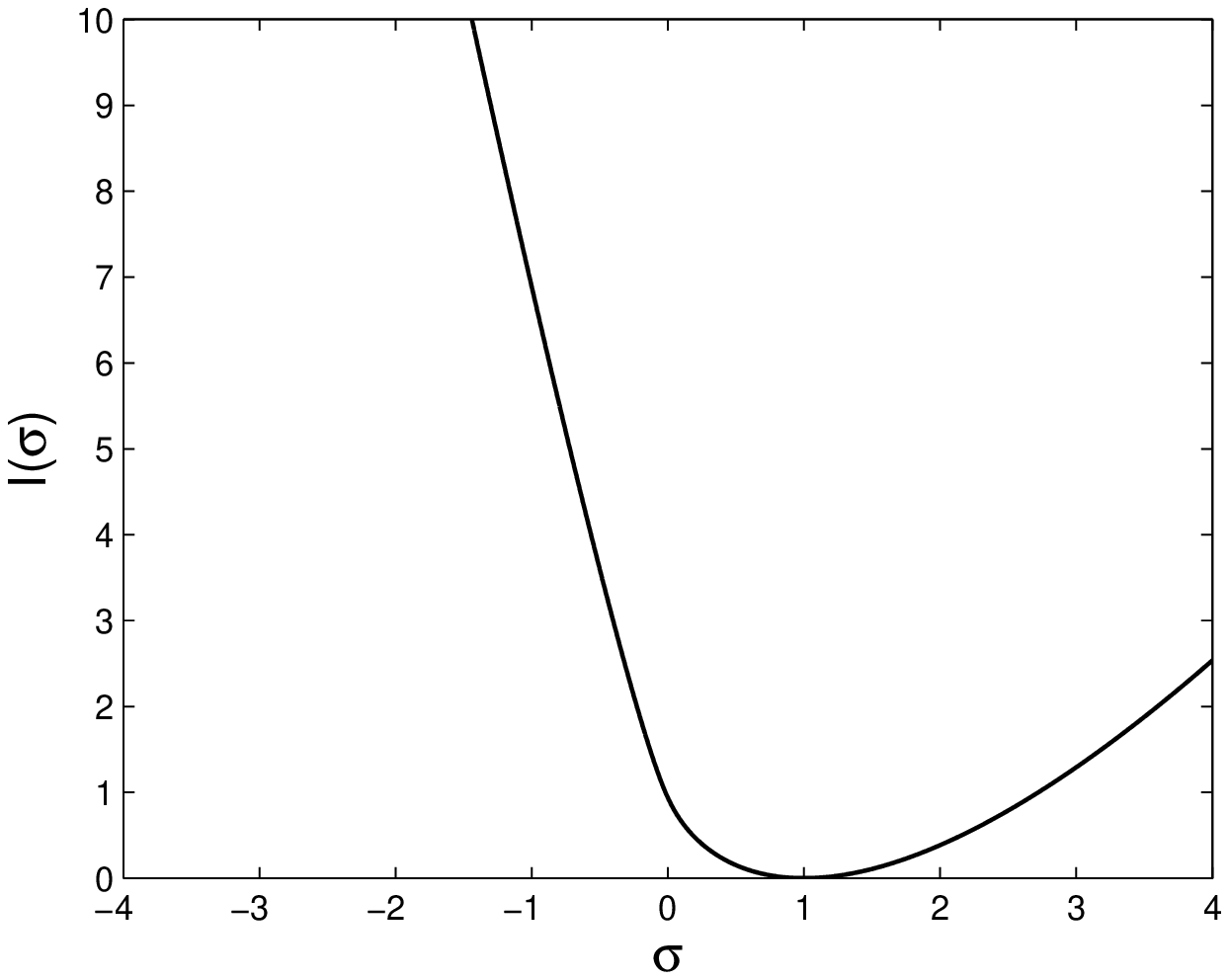}
(b)   \includegraphics[width=.45\textwidth,clip,
angle=0]{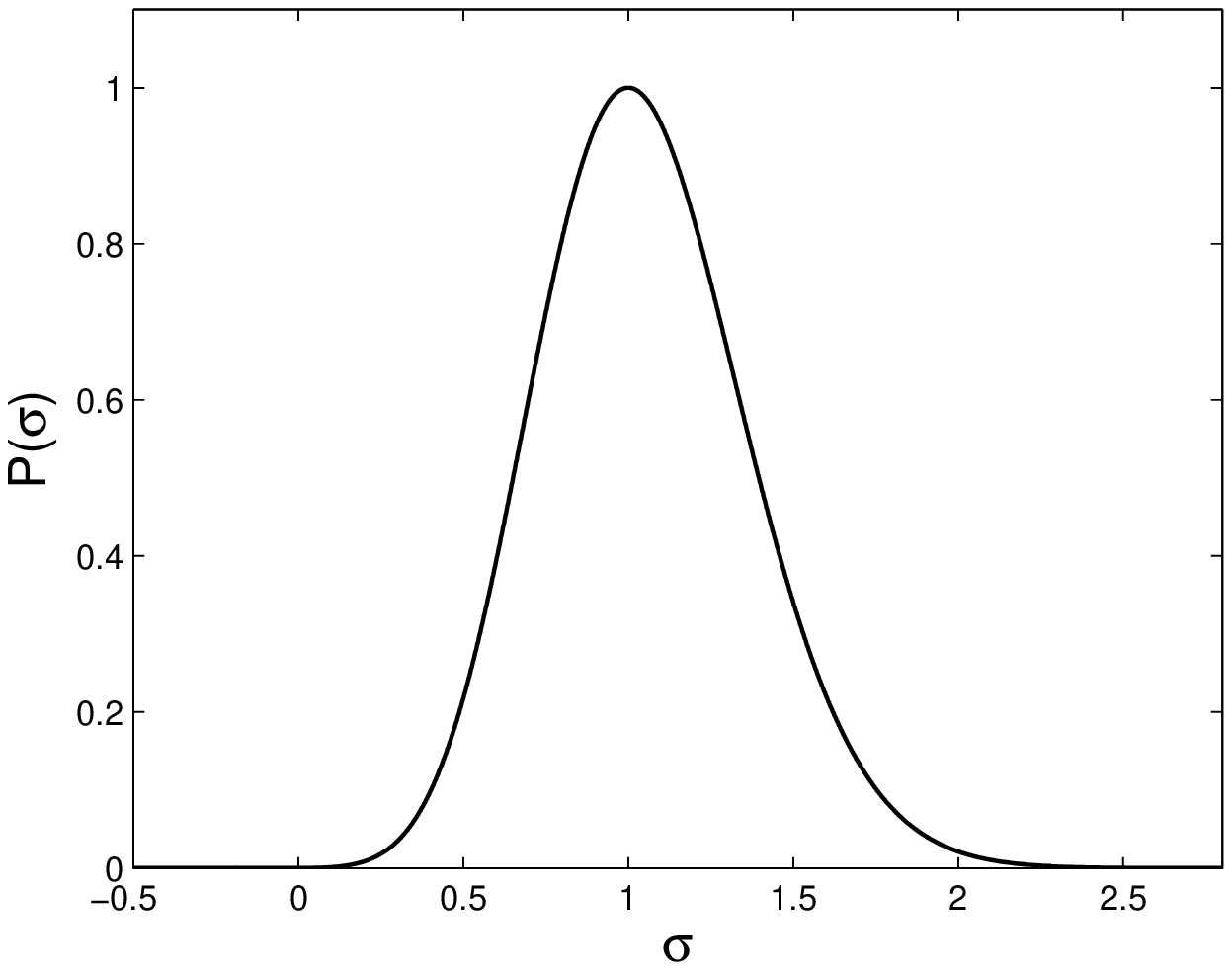}
\caption{(a) Large deviation function for the scaled entropy production $\sigma$ for  $\tau=0.001$. (b) Distribution of the scaled entropy production at time $t=10$.}
\label{fig:largedeviation_pleimling}
\end{figure}
The LDF for the normalized entropy production $\sigma=\Delta S_m/(t\langle \dot{S}_m\rangle)$ has the form
\begin{eqnarray}
I(\sigma)=\mu_0(\lambda^*)-\lambda^* \sigma\left\langle \dot{S}_m\right\rangle, \label{largedev_2}
\end{eqnarray}
where the saddle point $\lambda^*$ is defined implicitly as
\begin{equation}
\tau^{\lambda^*}(1-2\tau)^{1-\lambda^*}-(1-2\tau)^{\lambda^*}\tau^{1-\lambda^*}=\sigma(1-3\tau).\label{implicit_eq}
\end{equation}
 Expressing equation (\ref{implicit_eq}) in terms of $x$, where 
$x=\tau^{\lambda^*}(1-2\tau)^{-\lambda^*}$,  we  find 
\begin{equation}
x=\frac{\sigma(1-3\tau) +\sqrt{\sigma^2(1-3\tau)^2 +4\tau(1-2\tau)}}{2(1-2\tau)}. \label{tausol}
\end{equation}
The saddle point $\lambda^*$ is related to $x$ as
\begin{equation}
\lambda^* =-(1-3\tau)\frac{\ln(x)}{\langle\dot{S}_m\rangle}.\label{saddle_lambda}
\end{equation}
Substituting equation(\ref{smallesteigenv}), (\ref{med_entropy_pleimling}) and (\ref{saddle_lambda}) into (\ref{largedev_2}), we have
\begin{eqnarray}
I(\sigma)=1-\tau-\frac{\tau}{x}-(1-2\tau)x +\sigma(1-3\tau) \ln(x). \label{ld_pleimling}
\end{eqnarray}
The symmetry property of the LDF, $I(\sigma)-I(-\sigma)= -\langle\dot{S}_m\rangle \sigma$, implies that the fluctuation relation for the entropy production in the medium  holds for the system in the 
 long time limit.  The  plot of the LDF for the entropy production (see figure  \ref{fig:largedeviation_pleimling}(a))  shows a kink at zero entropy production as a consequence of the fluctuation theorem \cite{pleimling1,pleimling2}. The distribution function for the entropy 
 production, as shown in \ref{fig:largedeviation_pleimling}(b),  is  obtained directly from equation (\ref{ld_pleimling}).

\section{Summary and future perspectives}\label{sec:5}
In summary, we have obtained the LDF and the probability distribution function 
for the medium entropy production 
 for a two-site TASEP and a three-state cyclic process. Both TASEP and  the 
 three-state process involve 
 irreversible transitions due to the hopping of particles in a 
 specific direction and the unicyclic nature of the three-state process. 
 In order to apply the general results of entropy production 
 for stochastic jump processes, we obtained first the time-dependent transition 
 rates by sampling the states of the systems over a  short time interval. 
 These new  transition rates are incorporated in the subsequent derivations 
 of the LDF for the entropy production which 
 satisfies Gallavotti-Cohen symmetry. For the two-site TASEP, the value of the 
 LDF for large fluctuations becomes higher as the average entropy
  production rate is increased. As a consequence of this,  the distribution 
  function tends to be Gaussian. For smaller values of particle injection 
  and withdrawal rates, which in turn makes the average entropy
   production rate smaller, the distribution function becomes non-Gaussian 
   and it resembles Poisson distribution because of lesser number 
 of events over the time interval. For the three-state irreversible loop,
  we have found the  analytical forms of the smallest eigenvalue 
and the LDF. For both  the processes, the smallest 
eigenvalue and the LDF are derived  keeping the first order 
terms in $\tau$ in the  new transition rates. The 
smallest eigenvalue and the LDF for the medium entropy 
production satisfy the fluctuation theorem. Our results 
for the three-state 
process differ slightly from the previous 
study \cite{pleimling1}
 since we incorporate here the conventional definition of  the 
transition rates in the subsequent derivations of the 
smallest eigenvalue and the LDF. In reference \cite{ohkubo}, applying Bayes theorem to the posterior probabilities, it has been shown that the microscopic reversibility condition is not a necessity to propose a generalize fluctuation theorem for total entropy production. Since using time coarse-graining procedure, we obtain nonzero reverse transition probabilities even for processes involving irreversible transitions, it is expected that this procedure leads to holding  symmetry relations of certain kinds for the probability distribution of entropy production. Using similar coarsening theorem, the authors in reference \cite{pleimling3} have shown the validity of integral fluctuation theorem and Crooks relation for Hatano-Sasa entropy of many-state irreversible processes.  Finally, the present 
methodology based on derivation of  the 
time-dependent transition rates  seems to be useful 
for studying a broad range of models relevant to physical 
and  biophysical processes including complex networks involving multiple degrees of freedom \cite{pleimling3,robert}.

\end{document}